# Крутизна и спектр нелинейно деформируемой волны на мелководье

И.И.Диденкулова[1,2], Н.Заибо[3], А.А.Куркин[1], Е.Н.Пелиновский[1,2]

[1] Кафедра прикладной математики, Нижегородский государственный технический университет, Нижний Новгород

[2] Отдел нелинейных процессов в геофизике, Институт прикладной физики РАН, Нижний Новгород

[3] Кафедра физики, Университет Антильских островов, Пуэнт-а-Питр, Франция

Исследован процесс нелинейной деформации поверхностной волны на мелководье. Основное внимание уделено связи между Фурье-спектром волны и крутизной переднего склона. Показано, что в случае первоначально синусоидальной волны между ними существует однозначная связь, что позволяет оценивать спектральный состав волнового поля по наблюдаемой крутизне волны.

## Steepness and spectrum of nonlinear deformed wave in shallow water

Irina Didenkulova, Narcisse Zahibo, Andrey Kurkin, and Efim Pelinovsky

Process of the nonlinear deformation of the surface wave in shallow water is studied. Main attention is paid to the relation between the Fourier – spectrum and wave steepness. It is shown that the spectral harmonics of the initially sine wave can be expressed through the wave steepness, and this is important for applications.

Процесс нелинейной деформации волны на мелководье, приводящий к ее обрушению, хорошо известен, и в рамках нелинейной теории мелкой воды допускает точное аналитическое описание (см., например, Стокер, 1959: Вольцингер и др., 1989; Арсеньев и др., 1991). При этом основное внимание уделяют так называемой градиентной катастрофе (пересечению характеристик гиперболической системы мелкой воды), что должно соответствовать на практике обрушению переднего склона волны. Недавнее катастрофическое цунами в Индийском океане, случившееся 26 декабря 2004 года, во многих прибрежных пунктах проявилось накатом обрушенной волны, фактически «стеной воды». Демонстрирующие это видеофильмы сейчас легко доступны через Интернет. Тем не менее, опрокидывание волны чаще всего случается вблизи берега или при вхождении в устье реки (Пелиновский, 1982; Пелиновский и Трошина, 1993; Tsuji et al, 1991; Wu and Tian, 2000; Caputo and Stepanyants, 2003; Zahibo et al, 2006), и большую часть времени волна распространяется как необрушенная, при этом меняется ее крутизна и спектр. Исследование спектральных свойств и крутизны мелководной волны представляет практический интерес и здесь будет проведено для регулярных волн.

Основные уравнения нелинейной теории мелкой воды имеют вид

$$\frac{\partial u}{\partial t} + u\frac{\partial u}{\partial x} + g\frac{\partial \eta}{\partial x} = 0,$$

$$\frac{\partial \eta}{\partial t} + \frac{\partial}{\partial x}\left[(h+\eta)u\right] = 0,$$

(1)

где $\eta$ − возвышение водной поверхности, $u$ − горизонтальная скорость водного потока, $g$ − ускорение силы тяжести и $h$ − невозмущенная глубина бассейна, предполагаемая постоянной. Рассматривая волны, движущиеся в одну сторону (для определенности в сторону $x > 0$), порядок системы (1) может быть понижен (так как в волне существует однозначная связь между скоростью и смещением), и получаемое уравнение может быть записано в универсальной форме для мгновенной скорости распространения волны, $V$ (см., например, Вольцингер и др., 1989)

$$\frac{\partial V}{\partial t} + V\frac{\partial V}{\partial x} = 0,$$

(2)

$$V = \sqrt{gh} + \frac{3u}{2} = 3\sqrt{g(h+\eta)} - 2\sqrt{gh}, \qquad u = 2\left(\sqrt{g(h+\eta)} - \sqrt{gh}\right).$$

Подчеркнем, что уравнение (2) является точным при пренебрежении диссипацией и дисперсией, и справедливо для волны любой амплитуды вплоть до ее обрушения. Предполагая, что в начальный момент времени задана волна $\eta(x,t=0) = \eta_0(x)$, решение уравнения (2) имеет вид

$$V(x,t) = V_0(x - Vt), \tag{3}$$

где $V_0(x)$ легко вычисляется по заданной функции $\eta_0(x)$ с помощью алгебраических выражений в (2). Неявная форма решения (3) описывает так называемую волну Римана, хорошо известную в нелинейной акустике (Руденко и Солуян, 1975; Гурбатов и др., 1990). Применительно к волнам на воде решение (3) описывает нелинейную деформацию волны с укручением ее переднего склона. Крутизна волны легко находится дифференцированием (3) с использованием связей (2)

$$\frac{\partial \eta}{\partial x} = \frac{\eta_0'}{1 + t V_0'}, \tag{4}$$

где штрих означает производную по аргументу функции в (3). На переднем склоне волны ($\partial\eta/\partial x < 0$) производная $\partial V_0/\partial x$ также отрицательна, и знаменатель в (4) убывает со временем, так что крутизна волны увеличивается и в момент времени $t=T$ обращается в бесконечность. Момент первого обрушения (время нелинейности) равно

$$T = \frac{1}{\max(-V_0')}. \tag{5}$$

Отсюда следует, что волна начинается разрушаться в точке на профиле с максимальным перепадом скорости распространения, которая, вообще говоря, не соответствует точке с максимальной крутизной. В качестве конкретного примера рассмотрим начальную синусоидальную волну вида $\eta_0(x) = a\sin(kx)$ с максимальной крутизной $s_0 = ak$ в точке с нулевым смещением. Обрушение начинается на впадине волны, и фаза этой точки зависит от амплитуды (рис. 1)

$$kx_* = \arcsin\left(\frac{\sqrt{1-(a/h)^2}-1}{a/h}\right), \qquad (6)$$

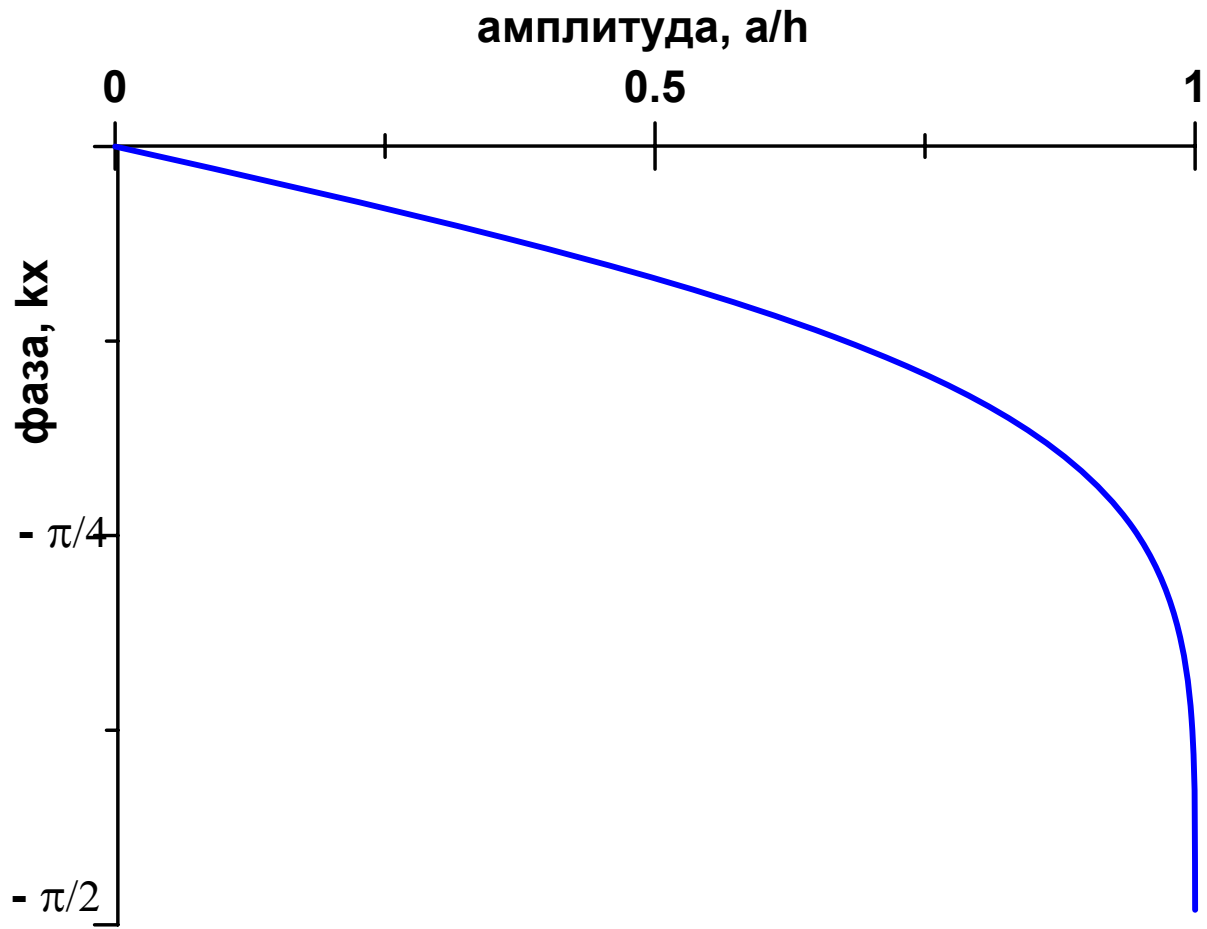

Рис. 1. Фаза волны, в которой начинается обрушение

Смещение уровня воды в этой точке есть (рис. 2)

$$\frac{\eta_*}{h} = \sqrt{1-(a/h)^2}-1. \qquad (7)$$

В волне малой амплитуды обрушение происходит почти на нулевом уровне, а в волне большой амплитуды обрушение начинается почти у дна.

Длина обрушения уменьшается с ростом амплитуды волны (рис. 3), стремясь к $X_{min} = (2)^{0.5}/3k$, и становится сравнимой с длиной волны. Таким образом, волна большой амплитуды опрокидывается фактически сразу в момент образования, что объясняется тем, что подножие волны касается дна. В то же время при малой амплитуде длина обрушения обратно пропорциональна амплитуде волны и может быть достаточно большой, так что волна остается необрушенной на большом расстоянии.

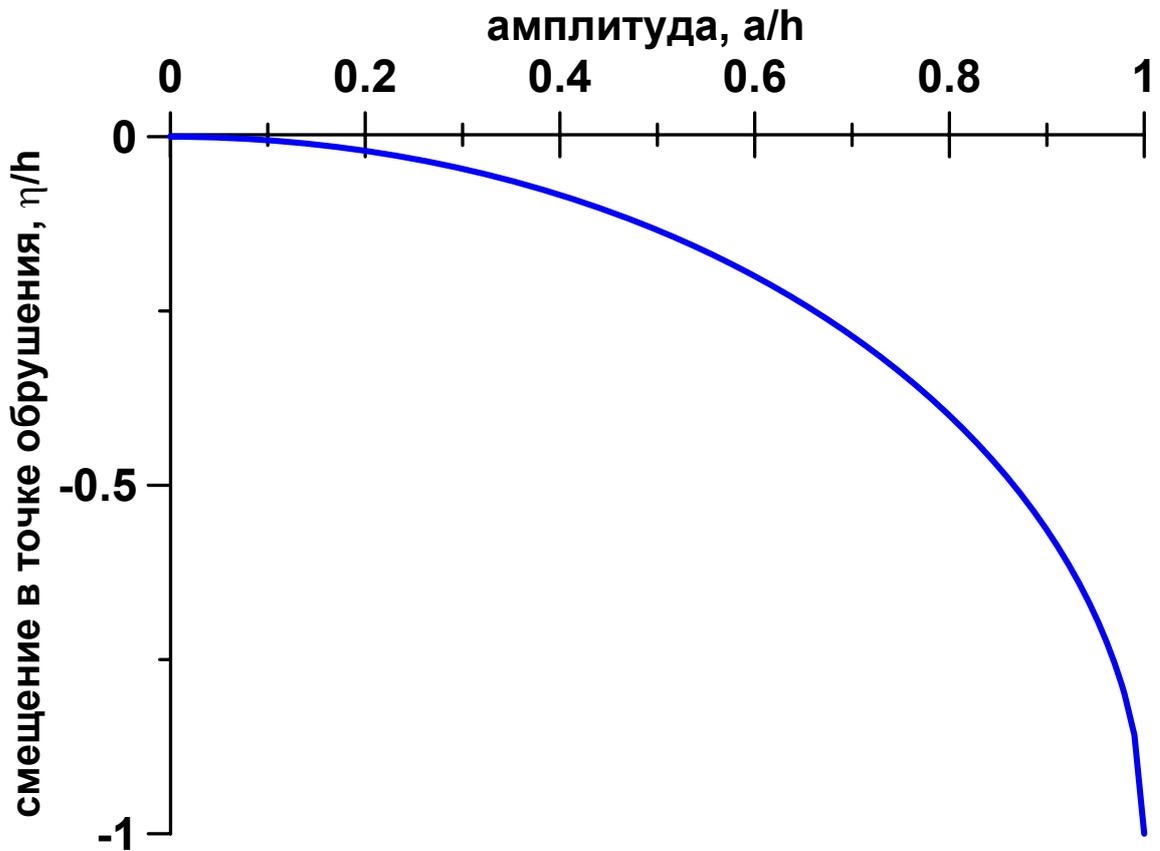

Рис. 2. Смещение уровня воды в точке обрушения

Время обрушения из (5) находится также в явном виде (удобно перейти к длине обрушения с помощью линейной скорости распространения длинных волн)

$$X = \sqrt{gh}T = \frac{1}{3k}\sqrt{\frac{2}{1-\sqrt{1-(a/h)^2}}}\ . \qquad (8)$$

И, наконец, из (4) находится простая зависимость максимальной крутизны переднего склона волны от времени

$$s = \max(\partial \eta / \partial x) = \frac{s_0}{1 - t/T}, \qquad (9)$$

где $s_0 = ak$, как уже говорилось, начальная крутизна волны.

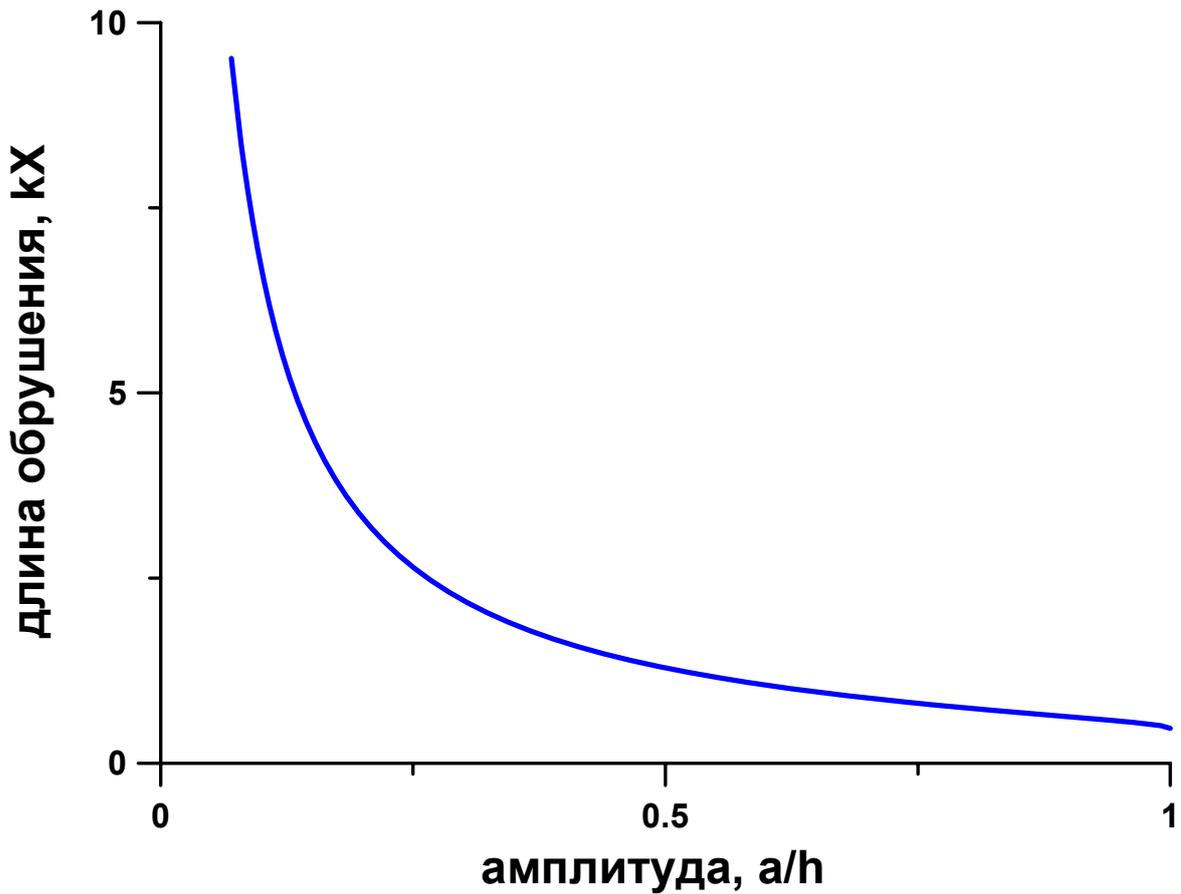

Рис. 3. Длина обрушения первоначально синусоидальной волны

Для практических приложений очень важно знать спектральный состав волнового поля. Пространственный спектр волны Римана в самом общем виде может быть записан в интегральной явной форме (Пелиновский, 1976)

$$S(k) = \int \eta(x,t)\exp(-ikx)dx = \frac{1}{ik}\int \frac{d\eta_0}{d\zeta}\exp\left(-ik[\zeta + tV(\eta_0)]\right)d\zeta . \qquad (10)$$

Вычислить этот интеграл даже для синусоидального начального возмущения в общем случае не удается аналитически. Рассмотрим поэтому здесь наиболее интересный случай волны малой, но конечной амплитуды, когда волна проходит большое расстояние прежде, чем обрушится. В этом случае воспользуемся разложением функции $V(\eta)$ в ряд Тейлора $V(\eta) = (gh)^{1/2}(1+3\eta/2h)$, тогда интеграл (10) вычисляется в явном виде и волновое поле в любой момент времени есть

$$\eta(t,x) = \sum_{n=1}^{\infty} A_n(t)\sin\left(nk[x - \sqrt{gh}\,t]\right) = \frac{4h}{3kt\sqrt{gh}}\sum_{n=1}^{\infty}\frac{1}{n}J_n\left(\frac{3nkta\sqrt{gh}}{2h}\right)\sin\left(nk[x - \sqrt{gh}\,t]\right), \quad (11)$$

где $J_n$ – функции Бесселя, а $n$ – номер гармоники. Вычисляя с той же точностью время обрушения из (8), спектральные амплитуды переписывается в виде

$$A_n(t) = 2a\frac{T}{nt}J_n\left(\frac{nt}{T}\right). \qquad (12)$$

Со временем амплитуды обертонов растут, а амплитуда основной гармоники падает, поскольку энергия волнового поля переходит вверх по спектру (рис. 4). Важно подчеркнуть, что даже в момент обрушения амплитуды обертонов относительно малы и быстро спадают с возрастанием номера гармоники.

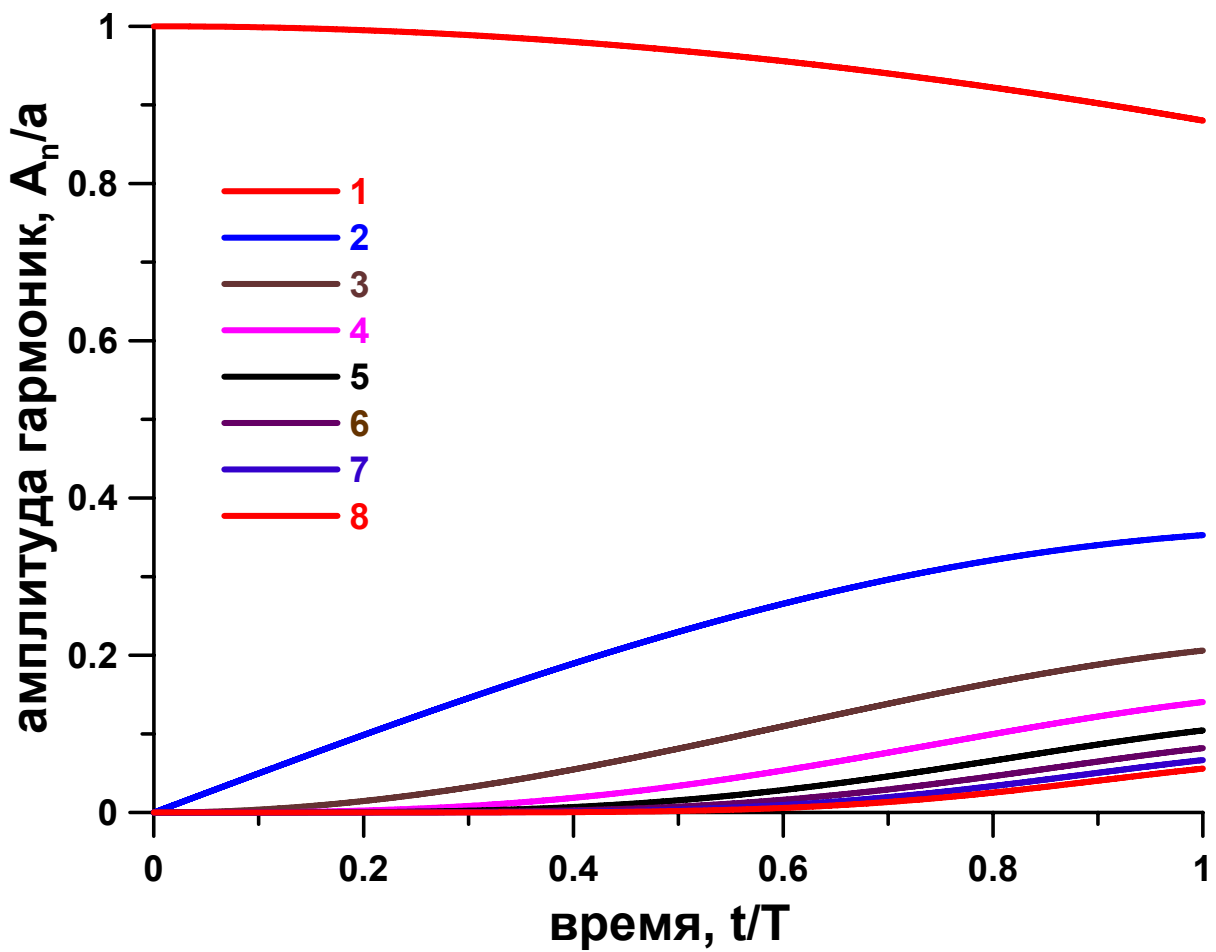

Рис. 4. Изменение амплитуд гармоник со временем (цифры – номера гармоник)

На практике расстояние от момента зарождения волны обычно неизвестно. Однако можно попробовать связать наблюдаемые характеристики волны, в частности, спектр и крутизну волны. Используя (9), запишем амплитуды гармоник в другом виде (рис. 5)

$$A_n(s) = \frac{2a}{n(1 - s_0/s)} J_n\left(n\left[1 - \frac{s_0}{s}\right]\right). \quad (13)$$

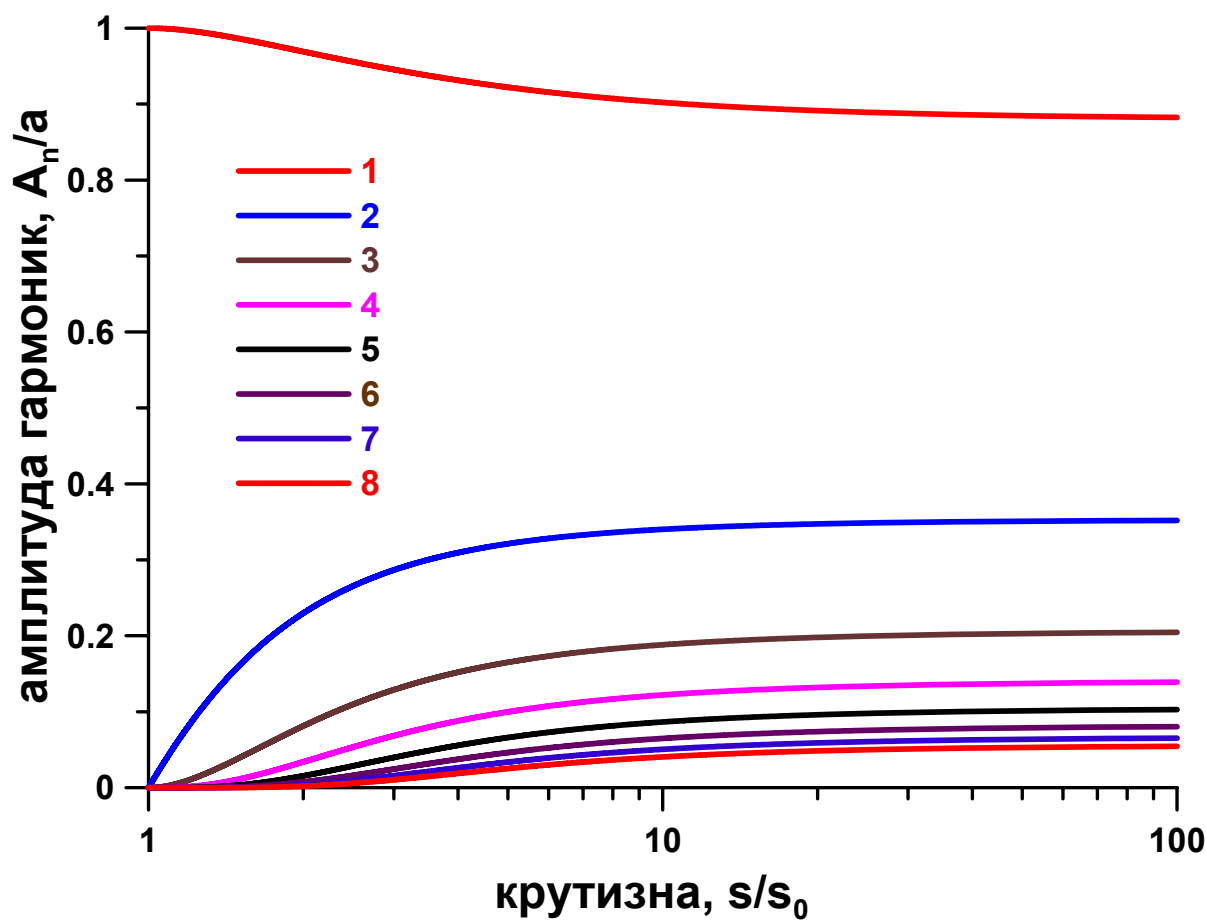

Рис. 5. Связь амплитуд гармоник с максимальной крутизной переднего склона волны

Как видно из рисунка, уже при увеличении крутизны переднего склона волны примерно на порядок, амплитуды гармоник выходят на постоянный уровень и перестают зависеть от крутизны волны. Предельный спектр мелководной волны не зависит от ее начальной крутизны

$$\overline{A}_n = \frac{2a}{n} J_n(n), \quad (14)$$

он показан на рис. 6. Эта функция очень хорошо аппроксимируется зависимостью $n^{-1.3}$, показанной на рисунке сплошной линией. В теоретическом плане спектр случайных римановых волн исследовался в нелинейной акустике (Гурбатов и др., 1990). Перед обрушением в спектре формируется асимптотика $k^{-3/2}$, что близко к рассчитанной нами для периодических волн. При дальнейшем распространении, когда обрушится уже весь гребень, в коротковолновой части спектра должна формироваться асимптотика $k^{-1}$, отражающая факт разрывности функции.

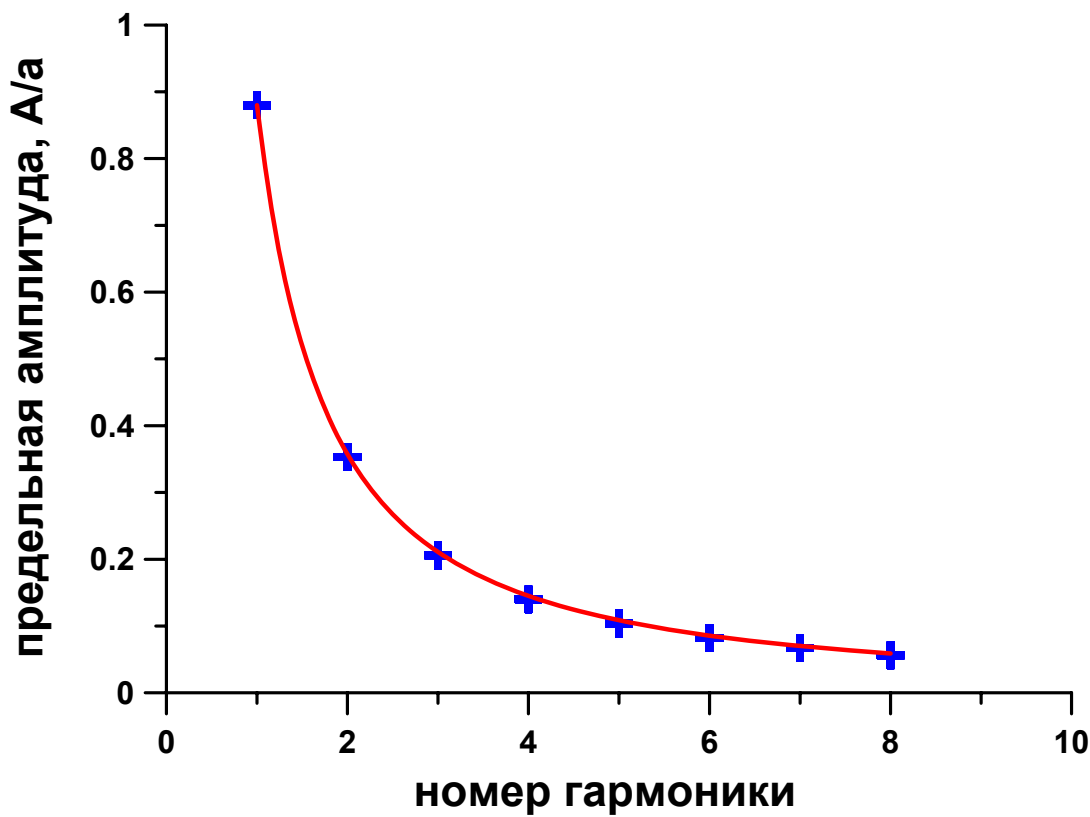

Рис. 6. Предельный спектр мелководной волны вблизи обрушения (сплошная линия – зависимость $n^{-1.3}$)

Итак, спектральные амплитуды нелинейной волны на мелководье можно оценивать по крутизне волны, причем, при значительной крутизне спектр становится универсальным. Это означает, что такие оценки являются достаточными грубыми и могут использоваться в практических расчетах.

**Литература**